\documentclass[aps,prl,reprint,twocolumn,showpacs,amsmath,amssymb,superscriptaddress]{revtex4}%
\usepackage{multirow}
\usepackage{bigstrut}
\usepackage{amssymb}
\usepackage{amsmath}
\usepackage{graphicx}
\usepackage{dcolumn}
\usepackage{bm}
\usepackage{CJK}
\usepackage{relsize}
\usepackage{cancel}
\usepackage{color}

\begin{document}

\title{Magnetic Field Modulated Resonant Tunneling in Ferromagnetic-Insulator-Nonmagnetic Junctions}

\author{Yang Song}\email{yangsong@pas.rochester.edu}\affiliation{Department of Electrical and Computer Engineering, University of Rochester, Rochester, New York, 14627}
\author{Hanan Dery}\affiliation{Department of Electrical and Computer Engineering, University of Rochester, Rochester, New York, 14627}\affiliation{Department of Physics and Astronomy, University of Rochester, Rochester, New York, 14627}

\begin{abstract}
We present a theory for resonance-tunneling magnetoresistance (MR) in Ferromagnetic-Insulator-Nonmagnetic junctions. The theory sheds light on many of the recent electrical spin injection experiments, suggesting that this MR effect rather than spin accumulation in the nonmagnetic channel corresponds to the electrically detected signal. We quantify the dependence of the tunnel current on the magnetic field by quantum rate equations derived from the Anderson impurity model, with important addition of impurity spin interactions. Considering the on-site Coulomb correlation, the MR effect is caused by competition between the field, spin interactions  and coupling to the magnetic lead. By extending the theory, we present a basis for operation of  novel nm-size memories.
\end{abstract}
\maketitle
Spintronics applications essentially rely on injection, manipulation and detection of spins \cite{Zutic_RMP04}. Demonstrations of electrically-injected spin accumulation in nonmagnetic materials are considered reliable when measured in a non-local geometry \cite{Johnson_PRB87,Jedema_Nature02}. In this setup, shown in Fig.~1(a), one ferromagnetic electrode injects or extracts spin-polarized electrons and a second one detects the spin accumulation of electrons ($V_{NL}$) that diffuse outside the path of a constant charge current ($I_T$). Because the spin diffusion length of many nonmagnetic materials is in the $\lesssim$~1~$\mu$m range, it is advantageous to have a submicron separation between the injector and detector electrodes \cite{Ji_APL06,Casanova_PRB09,Sasaki_APL11}. To mitigate this requirement, many researchers have recently resorted to a local measurement wherein one ferromagnetic electrode is used for both injection and detection of the spin signal [$V$ in Fig.~1(a)] \cite{Dash_Nature09,Tran_PRL09,Li_NatureComm11,Ando_APL11,Joen_APL11,Gray_APL11,Dash_PRB11,Jain_PRL12,
Erve_NatureNano12,Aoki_PRB12,Ishikawa_APL12, Uemura_APL12,Han_NatureComm13,Jeon_PRB13,Pu_APL13,Txoperena_APL13}.
Figure 1(b) shows the typically observed change in the detected resistance when applying an external magnetic field. Similar to the Hanle-type experiment of optical spin injection \cite{Hanle_1924,optical_orientation,Zutic_RMP04,Lou_PRL06}, the width and amplitude of the Lorentzian-shaped signal ($\Delta$$B$ and $\Delta R$) are frequently used to extract the spin lifetime and accumulation density. A critical problem, however, is that standard spin diffusion and relaxation theories cannot explain many of the recent local-setup experiments. Firstly, $\Delta$$B$ and $\Delta R$ are surprisingly insensitive to which nonmagnetic material is employed. Secondly, $\Delta R$ is too large to account for spin accumulation while $\Delta$$B$ is oddly comparable for electrons and holes. These facts raise big questions on the underlying physics, especially in technologically relevant materials such as silicon.

\begin{figure}[t]
\includegraphics[width=8.5cm]{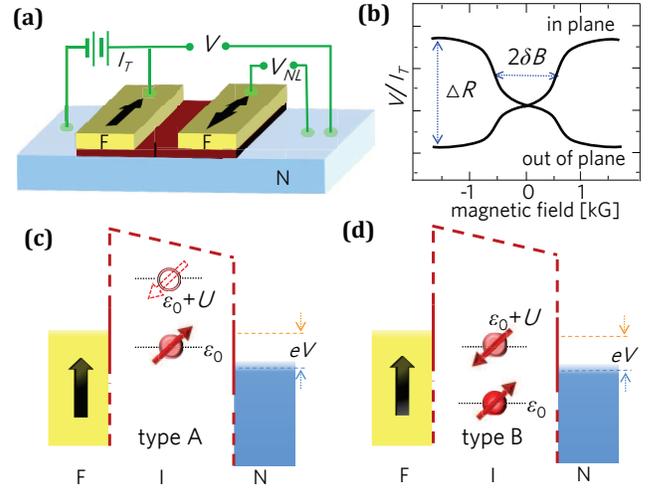}
\caption{(Color online) (a) Nonlocal and local electrical setups for detecting spin accumulation. (b) The measured signal, $\delta R(B) = [V(B)-V(0)]/I_T$, is a change in junction resistance when applying in-plane or out-of-plane magnetic fields. The Lorentzian due to in-plane field is typically observed only in the local setup. (c)/(d) Resonant tunneling via type A/B impurities for spin injection (electrons flow from F to N).} \label{fig:real_recip_lattice} \vspace{-3mm}
\end{figure}

In this Letter, we present a theory for resonance-tunneling magnetoresistance (MR) in Ferromagnet-Insulator-Nonmagnetic (F-I-N) junctions. We explain the greatly enhanced spin signals in numerous local spin injection/detection setups,   showing that $\Delta B$ and $\Delta R$ do not depend on spin accumulation and relaxation  in N. As the detector junction remains unbiased only for the nonlocal setup, the local detection is highly prone to impurity-assisted tunnel current. We propose that those enhanced signals and their dependence on temperature are set by impurities with large on-site Coulomb repulsion compared with the voltage bias. Depending on the electron occupation of the resonance level, the MR effect is established by the interplay between the Zeeman energy and the impurity coupling to F. Considering molecular fields due to spin-spin interactions such as hyperfine and exchange, we capture the origin of $\Delta B$ and the sign dependence of the signal on magnetic field orientation. Last but not least, by extending the theory to tunneling in one dimensional (1D) structures, we set forth a framework for a novel type of nanometer sized tunnel memory.

To quantify the MR, we tailor the Anderson impurity model to tunneling problem with spin polarized  leads \cite{Anderson_PR61,Appelbaum_PR69}. The energy of an electron with wavevector $\mathbf{k}$ and spin $\sigma$ in the $\ell$th-lead (F or N) is denoted $\varepsilon_{\ell\mathbf{k}\sigma}$. The creation (annihilation) Fermi operators in the lead and impurity are defined by $a_{\ell\mathbf{k}\sigma}^\dag(a_{\ell\mathbf{k}\sigma})$ and $d^{\dag}_{\sigma}(d_{\sigma})$, respectively.  The system Hamiltonian for $s$=$\tfrac{1}{2}$ impurity reads
\begin{eqnarray}\label{eq:Hamiltonian}
 H\!&\!=\!&\!
 \sum_{\ell\mathbf{k}\sigma} \!\left[ \varepsilon_{\ell\mathbf{k}\sigma} a_{\ell\mathbf{k}\sigma}^\dag a_{\ell\mathbf{k}\sigma}^{\,} + \! \left( T_{\ell\sigma}a_{\ell\mathbf{k}\sigma}^\dag d_{\sigma} + {\rm h.c.} \right)\right] \! + \! U n_{\uparrow} n_{\downarrow}
 \nonumber \\
 &&  +   \varepsilon_{\uparrow}(\theta) n_{\uparrow}  + \varepsilon_{\downarrow}(\theta) n_{\downarrow}  + \varepsilon_B  \sin\theta ( d^\dag_{\uparrow}d_{\downarrow}  +d^\dag_{\downarrow}d_{\uparrow}).
\end{eqnarray}
The interaction between the lead and impurity is denoted by $T_{\ell\sigma}$, assumed here to be $\mathbf{k}$ independent for simplicity. The on-site Coulomb interaction between electrons of opposite spins is denoted by $U$ and $n_{\sigma}=d^\dag_\sigma d_\sigma$. The $\sigma$$\,$=$\,$$\uparrow(\downarrow)$ component is parallel to the majority (minority) spin population of F. The second line in (\ref{eq:Hamiltonian}) denotes the impurity Zeeman terms where $\theta$ is the angle between $\mathbf{B}$ and the spin quantization direction. $\varepsilon_{\uparrow,\downarrow}$($\theta$)$\,$$=$$ \,$$\epsilon_0$$\,$$\pm$$\,$$\varepsilon_B$$\cos{\theta}$ where $\epsilon_0$ is the resonance energy of the singly occupied state and $\varepsilon_B = g\mu B/2$. The off-diagonal terms $d^{\dag}_{\sigma}d_{\bar{\sigma}}$ result from spin precession.

We briefly describe the derivation of resonance current.  The equation of motion for density-matrix operators is
\begin{eqnarray}\label{eq:EOM}
\!\!\!\!\!\!\! && \!\!\!\!\!\!\!  -i\hbar \frac{d}{dt}d^\dag_{\sigma}d_{\sigma'}  \equiv [H,d^\dag_{\sigma}d_{\sigma'} ] =  \sum_{\ell,\mathbf{k}} T_{\ell\sigma}a_{\ell\mathbf{k}\sigma}^\dag d_{\sigma'} - T_{\ell\sigma'}d_{\sigma}^{\dag}  a_{\ell\mathbf{k}\sigma'} \nonumber
\\ &&
+\,\, \varepsilon_B \sin{\theta}( d^\dag_{\bar\sigma}d_{\sigma'}  - d^\dag_{\sigma}d_{\bar{\sigma}'}  )  \pm 2\varepsilon_B \cos{\theta}   d^\dag_{\sigma}d_{\sigma'} \delta_{\bar\sigma\sigma'}.\qquad
\end{eqnarray}
Henceforth, the $+$/$-$ sign refers to the case that $\sigma$$\,$=$\,$$\uparrow$/$\downarrow$. To form a closed equation set, we use the Langreth theorem and recast the averages of the sum terms into lesser and retarded  Green functions on the impurity \cite{Langreth, Meir_PRL92, supple},
\begin{eqnarray}\label{eq:Langreth}
  \sum_{\ell,\mathbf{k}} T_{\ell\sigma} \langle a_{\ell\mathbf{k}\sigma}^\dag d_{\sigma'}  \rangle
  =
  \sum_{\ell}\! \int \!\! \frac{d\varepsilon}{2\pi} \Gamma_{\ell\sigma}\!\left( G^{R}_{\sigma'\sigma}f_{\ell\sigma} + \frac{1}{2} G^{<}_{\sigma'\sigma} \right).
\end{eqnarray}
$f_{\ell\sigma}(\varepsilon)$ is the Fermi distribution of $\sigma$ spin in the $\ell$th lead and $\Gamma_{\ell\sigma}(\varepsilon) \!=\! 2\pi \sum_{\mathbf{k}} |T_{\ell\sigma}|^2 \delta (\varepsilon-\varepsilon_{\ell\mathbf{k}\sigma})$ is its coupling to the impurity. The analysis is greatly simplified outside the Kondo regime  and by assuming weak coupling ($\Gamma\!\ll \! \{k_BT,eV \}$). We focus on two impurity types wherein the population of the resonance state fluctuates between zero and one [type A; see Fig.~1(c)] or between one and two electrons [type B; see Fig.~1(d)]. This classification is motivated by the dependence of the Green functions on the impurity population. It is justified when considering together the broad energy distribution of mid-gap impurity defects at oxide tunnel barriers \cite{Ephron_PRB94,Bahlouli_PRB94,Boero_PRL97,Lu_PRL02,Rippard_PRL02} and the large on-site Coulomb repulsion $U$. Under the common conditions of local-setup experiments $eV \gg k_BT \gg \varepsilon_B$,   we can replace $f_{\ell\sigma}(\varepsilon)$ by $1$ ($0$) for the injector (extractor) lead, and the Green functions take simple forms in Eq.~(\ref{eq:Langreth}) where $\Gamma(\varepsilon)$ varies slowly on the scale of $\varepsilon_B$ [e.g. $\int_{\scriptscriptstyle{e\!V}} d\varepsilon G^<_{\sigma'\sigma}= 2i\pi(\langle n_{\uparrow}\rangle+\langle n_{\downarrow}\rangle-1)\delta_{\sigma\sigma'}$ for type B; see supplemental material for details]. The analysis becomes independent of spin accumulation in the leads.  Putting these pieces together, we reach a concise equation set \cite{supple,Dong_PRB04,Braun_PRB04,Gurvitz_PRB96}. For spin extraction via type A impurities [electrons flow from N to F; opposite to the spin-injection bias setting in Fig.~1(c)],
\begin{eqnarray}\label{eq:rateA}
 \hbar \dot  n_{\sigma\sigma}  & = & \Gamma_{\rm N} P_0 - (1 \pm p)\Gamma_{\rm F} n_{\sigma\sigma} - 2 \varepsilon_B \sin{\theta} \text{Im}(n_{\bar{\sigma}\sigma}),
 \\
 \hbar \dot n_{\sigma\bar{\sigma}}  & = &  i\varepsilon_B\left[  \sin{\theta}( n_{\bar\sigma\bar\sigma} - n_{\sigma\sigma}) \pm 2\cos{\theta} n_{\sigma\bar{\sigma}} \right] -  \Gamma_{\rm F} n_{\sigma\bar{\sigma}}, \nonumber
\end{eqnarray}
where $ n_{\sigma\sigma'}\!\equiv \!\langle d^\dag_\sigma d_{\sigma'}\rangle$. $P_0\! =\! 1 \!-\! n_{\uparrow\uparrow} \!-\! n_{\downarrow\downarrow}$ is the probability for zero occupation and the coupling parameters $\Gamma_\ell(\varepsilon)\!=\!(\Gamma_{\ell\uparrow}\! +\! \Gamma_{\ell\downarrow})/2 $ are evaluated around the impurity's energy level $\epsilon_0$. Interface current polarization is given by $p\! =\!(\Gamma_{{\rm F}\uparrow}\!-\!\Gamma_{\rm F\downarrow})/2\Gamma_{\rm F}$ \cite{Zutic_RMP04}.  The master equations for injection conditions are obtained by exchanging $\Gamma_{\rm N\sigma}\! \equiv\! \Gamma_{\rm N}$ with $\Gamma_{\rm F\sigma}$. Similarly, the equations for type B impurities are obtained by evaluating $\Gamma_{\rm F(N)}$ around $ \epsilon_0 \!+\! U$, by considering double rather than zero occupancy ($P_2 = n_{\uparrow\uparrow} + n_{\downarrow\downarrow} - 1$), and by noting that type A and B impurities flip roles in extraction and injection conditions  \cite{supple}. This feature reflects their symmetry and can be viewed as  electron (hole) tunneling in type A (B) \cite{Glazman_JETPLet}.

The resonance currents are found from the steady state solution of the master equations using $i_{\rm A} =  2e\Gamma_{\rm N} P_{0}/\hbar$ and $i_{\rm B} =  e\Gamma_{\rm N} (1-P_{2})/\hbar$ for extraction, or $i_{\rm A} =  -e\Gamma_{\rm N} (1-P_{0})/\hbar$ and $i_{\rm B} =  -2e\Gamma_{\rm N} P_{2}/\hbar$ for injection [they implicitly relate to $\Gamma_{\rm F\sigma}n_{\sigma\sigma}$ by (\ref{eq:rateA})]. For $eV\gg k_B T$, we get \cite{supple}
\begin{eqnarray} \label{eq:curr}
&&\!\!\!\!\!\!\!\!\! i_{\rm A}^{\rm N \rightarrow \rm F}  =  -i_{\rm B}^{\rm F \rightarrow \rm N} =  \frac{2e}{\hbar} \frac{\Gamma_{\rm F} \Gamma_{\rm N}}{2\Gamma_{\rm N} + \Gamma_{\rm F}} \frac{1 - p^2 \chi(\mathbf{B})}{1 - \alpha p^2 \chi(\mathbf{B})}, \nonumber \\
&&\!\!\!\!\!\!\!\!\! i_{\rm B}^{\rm N \rightarrow \rm F}  =  -i_{\rm A}^{\rm F \rightarrow \rm N}   =  \frac{2e}{\hbar}\frac{\Gamma_{\rm F} \Gamma_{\rm N}}{2\Gamma_{\rm F} + \Gamma_{\rm N}}, \\
&&\!\!\!\!\!\!\!\!\! \chi(\mathbf{B})  =  \frac{B_{\rm F}^2 + B^2\!\cos^2{\theta}}{B_{\rm F}^2 + B^2}\,,\,\, \alpha =  \frac{\Gamma_{\rm F}}{ 2\Gamma_{\rm N}+ \Gamma_{\rm F}}\,,\,\,  B_{\rm F} = \frac{\Gamma_{\rm F}}{g\mu_B}. \nonumber
\end{eqnarray}
Most relevant to our analysis, the resonance current  across type A/B impurities depends on the magnetic field in extraction/injection conditions [via $\chi(\mathbf{B})$]. This dependence is best perceived when considering half-metallic F and out-of-plane magnetic field ($p=1$ and $\theta=\pi/2$). Without magnetic field, extraction via type A or injection via type B are completely blocked, $i_{\rm A}^{\rm N \rightarrow \rm F}$$\,$=$\,$$i_{\rm B}^{\rm F \rightarrow \rm N}$$\,$=$\,$0. In extraction via type A, electrons tunnel from N into the impurity and have equal probability to be parallel or antiparallel to the spin orientation in the half-metal. The tunnel conductance is blocked once an antiparallel spin settles on the impurity.  For injection via type B we get that once the lower impurity level is filled with an electron from the  half-metal, the upper resonant level can only accept the electron of opposite spin which the half metal cannot provide.  In a large out-of-plane field, the blockade is completely lifted in both cases due to depolarization of the impurity spin (Larmor precession). Finally, from (\ref{eq:curr}) we get that $i_{\rm A}$$\,$$+$$\,$$i_{\rm B}$ merely flips sign when reversing the bias direction. Therefore, the MR effect in injection (F$\rightarrow$N) and extraction (N$\rightarrow$F) is similar if the densities of type A and B impurities are similar.

To compare the analysis with experimental findings we incorporate important extensions on the effective magnetic field at the impurity site. When comprised of the external field alone, $\mathbf{B}$$\,$$=$$\,$$\mathbf{B}_{\rm e}$, the modulation amplitude $\Delta i(\theta_{\rm e}) \!= \! i(B_{\rm e} \! \gg \! B_{\rm F}\!) \!-\! i(B_{\rm F} \! \gg \! B_{\rm e}\!)$ is
\begin{eqnarray} \label{eq:deltaBF}
\Delta i(\theta_{\rm e})  = \frac{\sin^2{\!\theta_{\rm e}}}{1-\alpha p^2\cos^2{\theta_{\rm e}}} \frac{(1-\alpha)p^2}{1-\alpha p^2}  i_0\,\,,
\end{eqnarray}
where $i_0 = 2\alpha e \Gamma_{\rm N} /\hbar$. Throughout this work, $\mathbf{B}_{\rm e}$ is assumed smaller than the  out-of-plane coercive field of F. We see that the MR effect vanishes for in-plane external field ($\theta_{\rm e} = 0$) in contrast to most measurements where the in-plane field modulation is larger than that of the out-of-plane. Furthermore, for $\mathbf{B}$$\,$$=$$\,$$\mathbf{B}_{\rm e}$ the signal width stems from the coupling to F ($\Delta B \sim B_{\rm F}$), thereby decreasing exponentially with increasing oxide thickness. In virtually all local-setup measurements of F-I-N structures, on the other hand, $\Delta B \sim 0.1 - 1$~kG regardless of the oxide details \cite{Dash_Nature09,Tran_PRL09,Li_NatureComm11,Joen_APL11,Gray_APL11,Dash_PRB11,Jain_PRL12,Erve_NatureNano12,Aoki_PRB12,Uemura_APL12,Han_NatureComm13,Jeon_PRB13,Pu_APL13,Txoperena_APL13}.
To explain these aforementioned observations, we examine the ubiquitous spin interactions which tend to randomize the spin orientation at the impurities. They include, for example, hyperfine fields due to interaction with the nuclear spins and exchange interactions between nearby impurities. Invoking mean-field approximation, an effective internal magnetic field can be written by $\mathbf{B}_{\rm i} = \mathbf{B}_{\rm hf} + \mathbf{B}_{\rm ex} = (\langle A\mathbf{I}   \rangle + \langle J_{nn} \mathbf{S}   \rangle )/g\mu_B $ where $A$ is the hyperfine coupling constant with nuclear spin  $\mathbf{I}$ and $J_{nn}$ is the exchange coupling with an electron of spin $\mathbf{S}$ on the nearest neighbor impurity. Considering ferromagnet-oxide-silicon as a case study, unpaired electrons  on $\,$$^{29}$Si dangling bonds would experience hyperfine fields of a few hundred Gauss \cite{Griscom_PRB79,Jani_PRB83,Brower_APL83,Boero_PRL97,Lu_PRL02}. Similar defects can exist in Al$_2$O$_3$ barriers \cite{Rippard_PRL02,Stashans_PRB94,Momida_APL12,Choi_JAP13} or perovskite interfaces \cite{Pentcheva_PRB06}. The defect densities can be controlled by oxide preparation techniques \cite{Txoperena_APL13}. The sources for $\mathbf{B}_{\rm i}$ also include stray fields whose amplitude and direction depend on the interface roughness \cite{Dash_PRB11}.

\begin{figure}
\includegraphics[width=8.5cm]{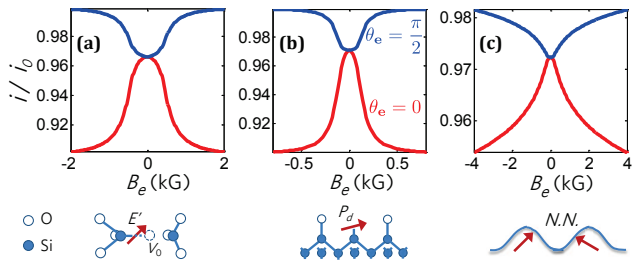}
\caption{(Color online) Calculated MR via an impurity with $\alpha=0.1$ embedded in the F-I-N junction with $p=1/3$. (a)/(b) MR due to resonant tunneling via an $E'$/$P_d$ defect in silicon-oxide interfaces. (c) MR in the presence of molecular fields due to exchange between nearest-neighbor impurities.} \label{fig:fig2} \vspace{-3mm}
\end{figure}

With $\mathbf{B}=\mathbf{B}_{\rm e} + \mathbf{B}_{\rm i}$, we can complete the analysis using (\ref{eq:curr}) and write the  extraction/injection tunnel current via a type A/B impurity, $i = i_{\rm A}= -i_{\rm B}$,
\begin{eqnarray} \label{eq:currBi}
\frac{i}{i_0}   \!=\!  1\! - \! (1\!-\!\alpha)p^2 \! \! \int \! \!\!dB_{\rm i} \!\! \int \!\!d\cos\!\theta_{\rm i} \!\int\!\! d\phi_{\rm i} \frac{ B^2_\|\mathcal{F}(B_{\rm i},\theta_{\rm i},\phi_{\rm i}) }{B_\bot^2\!\! +\! (1\!-\!\alpha p^2)B_\|^2}.
\end{eqnarray}
$i$ is averaged over $\mathcal{F}(B_{\rm i},\theta_{\rm i},\phi_{\rm i})$, the normalized distribution for the internal field. The components of the effective field along and normal to F are $B^2_\|=B_{\rm F}^2+(B_{i\|}+B_{e\|})^2$ and $B^2_\bot=B_{\rm i\bot}^2+B_{\rm e\bot}^2+2B_{\rm i\bot} B_{\rm e\bot} \cos(\phi_{\rm i}-\phi_{\rm e})$, respectively.
Figures~2(a) and (b) show the solution of (\ref{eq:currBi}) with hyperfine fields of common defect centers in Si/Oxide interfaces. The tunneling involves unpaired electrons  on $^{29}$Si dangling bonds next to oxygen vacancy $V_0$ in the barrier ($E'$ center) or in Si$_3$ configuration on the atomic interface ($P_d$ center). The hyperfine field of $E'$ is assumed isotropic with amplitude of 420~G \cite{Griscom_PRB79}, and that of $P_d$ has axial symmetry with an out-of-plane (in-plane) amplitude of 160~G  (90~G) \cite{Brower_APL83}. Figure~2(c) shows the solution for internal fields due to exchange between nearest-neighbor impurities. The localization length and impurity density in the tunnel barrier are chosen $\ell_i=4.4$~$\AA$ and $n_i=8\times10^{18}$~cm$^{-3}$, respectively. Further details are provided in the supplemental material \cite{supple}. In all three cases we assume $\langle B_{\rm i} \rangle > B_{\rm F}$ so that the width of the signal is set by internal fields rather than by coupling with F (i.e., $\Delta B$ is essentially independent of barrier thickness). The modulation amplitude in the regime that $B_{\rm i} \gg B_{\rm F}$ is realized from $\Delta i(\theta_{\rm e}) = i(B_{\rm e} \! \gg \! B_{\rm i}\!) \!-\! i(B_{\rm i} \! \gg \! B_{\rm e}\!)$. For isotropic internal field distribution [$\mathcal{F}(B_{\rm i},\theta_{\rm i},\phi_{\rm i})= \mathcal{F}(B_{\rm i})/4\pi$],
\begin{eqnarray} \label{eq:deltaBi}
\Delta i(\theta_{\rm e}) = \frac{1-\alpha}{\alpha} \left[  \frac{\text{arctanh}(\sqrt{\alpha} p)}{\sqrt{\alpha} p} -  \frac{1}{1-\alpha p^2\cos^2{\theta_{\rm e}}}\right] \!\!i_0.\,\,
\end{eqnarray}
The in-plane field  modulation  can exceed that of the out-of-plane and increases for internal fields that point mostly in the out-of-plane direction [e.g. $|\Delta i(0)/\Delta i(\pi/2)|>n+2$  when $\mathcal{F} \propto \sin^n{\!\theta_{\rm i}}\,\mathcal{F}(B_{\rm i})$].

We can now quantify the total voltage change that one measures in the local geometry [$\Delta R$  in Fig.~1(b)]. Denoting the total tunneling current used in experiments [Fig.~1(a)] as $I_T$, for small MR effect we simply have
\begin{eqnarray}\label{eq:dV}
\frac{\Delta R}{R} \!=\! \sum_{n}\frac{ \Delta i_n(0) \!-\! \Delta i_n(\frac{\pi}{2})}{I_T} = \frac{1}{I_T}\sum_{n}  \frac{(1\!-\! \alpha_n) p^2}{1 \!- \!\alpha_n p^2} i_{0,n} , \,\,\,
\end{eqnarray}
where $n$ runs over type A (B) impurities in the tunnel barrier for spin extraction (injection). The much larger total current is by tunneling via larger impurity clusters of which $U\!\lesssim \! eV$ and a background direct tunneling.  The MR effect is enabled by the nonzero polarization of a F-I-N junction ($p$$\,$$\neq$$\,$$0$), rendering it distinct from  resonant tunneling MR in N-I-N junctions where $g\mu B\!>\!\{k_B T, eV\}$ \cite{Glazman_JETPLet,Ephron_PRB94}.
Since $\Delta R$ measures the effect at $B_{\rm e} \!\gg\! \{B_{\rm i}, B_{\rm F}\}$ limit, its amplitude is robust and independent of the details of the internal field distribution. Accordingly, one can use either (\ref{eq:deltaBF}) or (\ref{eq:deltaBi}) to get (\ref{eq:dV}). The amplitude of $\Delta\! R/R$ depends on the junction's polarization, impurities density and their coupling to the leads (via $\alpha$ and $i_0$).


\textit{Discussion}. The MR effect in F-I-N junctions comes from electron spin precession in impurities whose population fluctuates between zero and one (type A)  when electrons flow into F, or between one and two (type B) when electrons flow from F. The resonance current through these impurities is suppressed or enabled when applying in-plane or out-of-plane magnetic fields, respectively. The physics is explained by reinforcement or removal of the Pauli-blockade in the respective field configurations.

The MR effect is stronger for impurities located closer to N than to F ($\Gamma_{\rm N} \!\gg \!\Gamma_{\rm F}$ so $\alpha \! \rightarrow  \! 0$). This physics is understood by noting that when electrons flow into (from) F, type A (B) impurities are mostly empty (doubly occupied) if they are closer to F. Therefore, spin precession becomes meaningless and the modulation is not observed. The disappearance of the effect for $\alpha \! \rightarrow \! 1$ also explains the results in a recent comprehensive experimental analysis of F-I-semiconductor junctions \cite{Sharma_PRB14}. A strong suppression in the MR signal is found when the oxide thickness decreases (exponential increase of $\Gamma_{\rm F}$), unlike the total $R$ that for ultrathin oxides is governed by the Schottky barrier ($\Gamma_{\rm N}$). This physics also sets apart the measurements of F-I-semiconductor junctions from those with \textit{direct} F-semiconductor contacts \cite{Ando_APL11,Lou_PRL06,Chan_PRB09, supple}. In the latter case, the true signal from spin accumulation in N cannot be  masked by the presence of impurities at the atomic interface between F and the semiconductor. The reason is that $\alpha \! \rightarrow \! 1$ for these impurities ($\Gamma_{\rm F} \!\gg \!\Gamma_{\rm N}$).

Thus far we have treated the on-site Coulomb repulsion as the largest energy scale  $U\gg eV$ (when the MR is most effective). Now we invoke the relation between $U$ and various sizes of impurity clusters in order to explain the nontrivial bias ($V$) and temperature ($T$) dependencies of the MR signal. We note that $U$ is smaller for relatively large clusters due to their reduced charging capacitance, and that the effective size of a cluster grows with $T$ due to the thermally activated crosstalk between adjacent
impurities \cite{Helman_PRL76}. $T$ dependence typically follows the Arrheniuss law with an activation energy $E_a$ that depends on disorder density and impurity type \cite{Txoperena_arxiv14}. Thus, as $k_B T$ rises above the corresponding $E_a$ such that for the resulting impurity cluster $U\lesssim eV$, this particular cluster stops affecting the MR. This interplay between $U$ and $eV$, and between $k_B T$ and $E_a$ resolves the strong dependence of $\Delta R(T)$ signals found in several recent  experiments \cite{Tran_PRL09,Sasaki_APL11,Jain_PRL12, Dash_Nature09,Pu_APL13}. At small bias $V$, the relevant $E_a$ for threshold $U_{\rm th} \approx eV$ is small as it corresponds to large and dense clusters, and as a result the MR effect is more susceptible to temperature at the low $T$ region $k_B T \approx E_a$. At large bias, on the other hand, the relevant $E_a$ becomes larger as the MR is from the outset limited to isolated point defects (largest $U$), for which the $T$ dependence is weaker ($k_B T \ll E_a$).

The proposed analysis solves two additional important problems in electrical spin injection. First, it addresses the observed signals in local-setup experiments where the net charge current across the tunnel junction is zero but where  the bias voltage  is distributed (i.e., spin injection and extraction in different parts of the junction) \cite{Breton_Nat11}. So far, the measured signals in such experimental settings were attributed to the spin Seebeck effect in spite of a similar orders-of-magnitude discrepancy with the theory of spin injection \cite{supple}. Second, the proposed mechanism supports the fact that the measured MR effect is independent of doping type in F-I-semiconductor junctions \cite{Dash_Nature09,Gray_APL11,Pu_APL13}. We have seen that the expected signal does not depend on spin relaxation in N, and therefore a comparable effect is expected in both \textit{n} and \textit{p}-type semiconductors. The original attribution to spin relaxation in N, on the other hand, contradicts known physics of ultrashort (sub-ps) spin lifetime in hole bands \cite{Hilton_PRL02, Loren_PRB11, Pezzoli_PRL12, supple}.

\begin{figure}
\includegraphics[width=8.5cm]{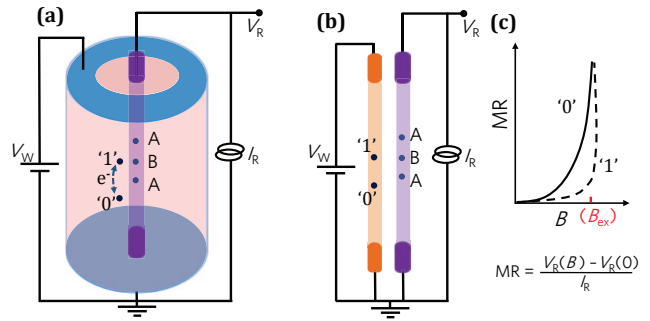}
\caption{(Color online) Nano-size tunnel memory relying on the MR effect. (a) The basic cell is a 1D conducting nanowire in an insulator (e.g., DXP molecules in zeolites \cite{Mahato_Science13}), or adjacent insulating and conducting wires as shown in (b). In either configuration, the writing voltage ($V_{\rm W}$) sets the position of an unpaired electron in one of two impurities embedded in the insulator (labeled `1' and `0').  The conducting wire includes a critical A-B-A impurity chain which becomes Pauli blocked when applying a magnetic field \cite{Txoperena_arxiv14}. (c) The MR effect facilitates the information readout due to its strong dependence on the exchange field of the unpaired electron (see text). } \label{fig:fig3} \vspace{-3mm}
\end{figure}

\textit{Outlook}. The MR mechanism can be generalized beyond spin injection with ferromagnetic leads. Figure~\ref{fig:fig3} shows such an example for a 1D nm-size memory cell that utilizes A-B impurity chains. Recent measurements in N-I-N tunnel junctions show that A-B chains result in a similar MR effect, where the type A impurity serves as an effective one-electron source with its polarization susceptible to weak  magnetic fields \cite{Txoperena_arxiv14}. A sufficient external field turns off the current by
reinforcing Pauli blockade across the A-B chain \cite{Txoperena_arxiv14}. As shown in Figs.~\ref{fig:fig3}(a) and (b), the `0' and `1' states are defined by the position of an unpaired electron embedded in an insulator adjacent to the A-B chain. Its position is controlled by the writing voltage $V_{\rm W}$.  The exchange interaction with the embedded electron when positioned in the `1' state sets the effective internal magnetic field ($B_{\rm ex}$) exerted on the type B impurity. The read-out is enabled by the MR effect across the A-B chain as shown in Fig.~\ref{fig:fig3}(c).  Note that confinement of the applied magnetic field is not needed since the spin does not encode information. Once the challenge  for atomic-level lithographic control is met, this architecture  represents the ultimate scaling of memories since `it leaves no room in the bottom'.

We are indebted to Felix Casanova, Oihana Txoperena, Kohei Hamaya, and Ian Appelbaum for insightful discussions and for sharing invaluable data prior to their publication. This work is supported by NRI-NSF, NSF, and DTRA Contract numbers DMR-1124601, ECCS-1231570, and HDTRA1-13-1-0013, respectively.

\end{document}


\title{Supplemental material for ``Magnetic Field Modulated Resonant Tunneling in Ferromagnetic-Insulator-Nonmagnetic (F-I-N) Junctions''}

\author{Yang Song}\email{yangsong@pas.rochester.edu}\affiliation{Department of Electrical and Computer Engineering, University of Rochester, Rochester, New York, 14627}
\author{Hanan Dery}\affiliation{Department of Electrical and Computer Engineering, University of Rochester, Rochester, New York, 14627}\affiliation{Department of Physics and Astronomy, University of Rochester, Rochester, New York, 14627}

\maketitle
\noindent {\bf 1. Derivation of Eq.~(3) in the main text.} From Eq.~(2), we readily obtain the kinetic equations for the average ensemble (`$\langle\rangle$') quantities of
\begin{eqnarray}
\frac{d }{d t}\langle n_{\sigma}\rangle
&=&
-\frac{2}{\hbar} \varepsilon_B \sin\theta {\rm Im}\langle d^\dag_{\bar{\sigma}} d_\sigma \rangle
 -\frac{2}{\hbar}\sum_{\ell,\mathbf{k}} {\rm Im} (T_{\ell\sigma} \langle a_{\ell\mathbf{k}\sigma}^\dag d_\sigma\rangle ),
 \label{eq:EOM_nd}
\\
\frac{d }{d t}\langle d^\dag_{\bar\sigma} d_\sigma \rangle
 &=&
\frac{i}{\hbar}  \varepsilon_B [2\bar{\sigma}\cos\theta \langle d^\dag_{\bar{\sigma}} d_\sigma  \rangle
+ \sin\theta (\langle n_{\sigma}\rangle- \langle n_{\bar{\sigma}} \rangle)]
+ \frac{i}{\hbar}\sum_{\ell,\mathbf{k}} ( T_{\ell\bar\sigma} \langle a_{\ell\mathbf{k}\bar\sigma}^\dag  d_\sigma \rangle- T^*_{\ell\sigma}\langle d^\dag_{\bar\sigma} a_{\ell\mathbf{k}\sigma}\rangle ).
\label{eq:EOM_dupdown}
\end{eqnarray}
$\sigma=\{+1,-1\}$ correspond, respectively, to spin $\{\uparrow,\downarrow\}$ in the algebra manipulation.
Next we make use of the so-called Langreth theorem to relate $ \langle a_{\ell\mathbf{k}\sigma}^\dag d_{\sigma'}\rangle$ with Fermi distribution in the leads ($  f_{\ell\mathbf{k}\sigma} = \langle a_{\ell\mathbf{k}\sigma}^\dag a_{\ell\mathbf{k}\sigma} \rangle $) and $\langle d^\dag_\sigma d_{\sigma'}\rangle$ of the impurity, so that we can have a closed set of equations for the impurity density matrix. By definition (under steady state),
\begin{eqnarray}\label{eq:lesser_green_kd}
\langle a_{\ell\mathbf{k}\sigma}^\dag d_{\sigma'}\rangle
=\langle a_{\ell\mathbf{k}\sigma}^\dag (t) d_{\sigma'} (t)\rangle
=\frac{1}{2\pi}\int d\varepsilon \langle a_{\ell\mathbf{k}\sigma}^\dag  d_{\sigma'} \rangle (\varepsilon)
=-\frac{i}{2\pi}\int d\varepsilon G^<_{\sigma', \ell \mathbf{k}\sigma } (\varepsilon).
\end{eqnarray}
By Langreth theorem \cite{Langreth_76, Meir_PRL92}, we have for the lesser Green function (GF)
\begin{eqnarray}\label{eq:Langreth_theorem}
G^<_{ \sigma', \ell \mathbf{k}\sigma} (\varepsilon)
= T^*_{\ell\sigma} \left[ G^R_{\sigma', \sigma}(\varepsilon) G^<_{\ell \mathbf{k}\sigma',  \ell \mathbf{k}\sigma}(\varepsilon)
+ G^<_{\sigma', \sigma} (\varepsilon) G^A_{\ell \mathbf{k}\sigma',  \ell \mathbf{k}\sigma} (\varepsilon)
\right],
\end{eqnarray}
where $G^{R(A)}$ denotes the usual retarded (advanced) GF.
Equations~(\ref{eq:lesser_green_kd}) and (\ref{eq:Langreth_theorem}) can be used to develop the last term in Eqs.~(\ref{eq:EOM_nd}) and (\ref{eq:EOM_dupdown}), which represents the current between the impurity and the leads,
\begin{eqnarray} \label{eq:current_Langreth}
\sum_{\mathbf{k}}  T_{\ell\sigma} \langle a^\dag_{\ell\mathbf{k}\sigma} d_{\sigma'} \rangle
&=&
\frac{-i}{2\pi}\int d\varepsilon \sum_{\mathbf{k}}  | T_{\ell\sigma}|^2
\left[
G^R_{\sigma', \sigma}(\varepsilon) G^<_{\ell \mathbf{k}\sigma',  \ell \mathbf{k}\sigma}(\varepsilon)
+ G^<_{\sigma', \sigma} (\varepsilon) G^A_{\ell \mathbf{k}\sigma',  \ell \mathbf{k}\sigma} (\varepsilon)
\right]
\nonumber\\
&=&
\int d\varepsilon \sum_{\mathbf{k}}  | T_{\ell\sigma}|^2
 \left[
 G^R_{\sigma', \sigma}(\varepsilon) f^{FD}_\ell(\varepsilon) \delta( \varepsilon-\varepsilon_{\ell \mathbf{k} \sigma })
+  \tfrac{1}{2}  G^<_{\sigma', \sigma} (\varepsilon) \delta( \varepsilon-\varepsilon_{\ell\mathbf{k}\sigma})
\right]
\nonumber\\
&=&
\frac{1}{2\pi}\int d\varepsilon  \Gamma_{\ell \sigma}(\varepsilon)
 \left[
 G^R_{\sigma', \sigma}(\varepsilon)  f^{FD}_\ell  (\varepsilon)
+  \tfrac{1}{2}  G^<_{\sigma', \sigma} (\varepsilon)
\right].
\end{eqnarray}

\noindent{\bf 2. Derivation of Eqs.~(4) in the main text.} The result of Eq.~(\ref{eq:current_Langreth}) can be further simplified for type A or type B regime. In the following, we give the detailed results of impurity GFs using the familiar equation-of-motion (EOM) technique \cite{Appelbaum_PR69,Lacroix_JPF81,Zubarev_SPU60}. Reduction to Eqs.~(4) for type A case and the counterpart equations for type B case can then be explicitly seen from these detailed results.
Briefly, the EOM for an energy-domain retarded GF, $\langle\langle A,B^\dag\rangle\rangle\equiv G^R_{A,B}$, is developed by
\begin{eqnarray}\label{eq:green_EOM}
\varepsilon \langle\langle A,B^\dag\rangle\rangle = \langle [A,B^\dag]\rangle +\langle\langle AH-HA,B^\dag\rangle\rangle
\end{eqnarray}
where $[A,B^\dag]=AB^\dag+(-)B^\dag A$ for fermion (boson) operators. The EOMs for $\langle \langle d_{\sigma}, d^\dag_{\sigma'} \rangle\rangle\equiv G^R_{\sigma\sigma'}$ generate higher order GFs,  $\langle \langle n_{-\sigma} d_{\sigma}, d^\dag_{\sigma'} \rangle\rangle$, which in turn are expressed with GFs of four particle operators. Truncation at this order, where only one operator refers to the leads, inevitably breaks the correlation at the impurity. We continue to the next order with ten GFs and their EOMs, and decouple in the usual way the resulting GFs which contain two operators for lead electrons. It is noteworthy to mention that  far above Kondo temperature, it is appropriate to use $\langle a_{\ell\mathbf{k}\sigma}^\dag  d_{\sigma'}\rangle \simeq 0$ and $\langle a_{\ell\mathbf{k}\sigma}^\dag a_{\ell\mathbf{k}'\sigma'} \rangle =  f_{\ell\mathbf{k}\sigma} \delta_{\mathbf{k},\mathbf{k}'}\delta_{\sigma,\sigma'}$. That is, the charge fluctuation and the correlation within the leads are neglected. Careful but straightforward derivation leads to two coupled equations of $\langle\langle d_\uparrow, d^\dag_\uparrow \rangle\rangle$ and $\langle\langle d_\downarrow, d^\dag_\uparrow \rangle\rangle$,
\begin{eqnarray}\label{eq:green_duu_ddu1}
&&\left[ \varepsilon -\varepsilon_\uparrow +i \Gamma_\uparrow
-i ( \varepsilon-\mathcal{E}^U_{\downarrow})\widetilde{\Gamma}_{\downarrow} \frac
{ U }{ \Lambda}\right]
G^R_{\uparrow \uparrow}
-
\varepsilon^B_{\uparrow\downarrow}\left(1
+i \widetilde{\Gamma}_{\uparrow}
\frac
{ U }{ \Lambda}\right)
G^R_{\downarrow \uparrow}
= 1
+
 \left[ n_{\downarrow\downarrow} ( \varepsilon-\mathcal{E}^U_{\downarrow} )
- n_{\uparrow\downarrow}  \varepsilon^B_{\uparrow\downarrow}\right]
\frac{ U }{ \Lambda}
\end{eqnarray}
and
\begin{eqnarray}\label{eq:green_duu_ddu2}
&& \left[\varepsilon -\varepsilon_\downarrow +i \Gamma_\downarrow
-
 i ( \varepsilon-\mathcal{E}^U_{\uparrow} )\widetilde{\Gamma}_{\uparrow}
 \frac{ U }{ \Lambda}\right]
G^R_{\downarrow\uparrow}
-
\varepsilon^B_{\uparrow\downarrow}\left(1
+
i \widetilde{\Gamma}_{\downarrow}
\frac{ U }{ \Lambda}
 \right)
G^R_{\uparrow\uparrow}
=
\left[
  n_{\downarrow\downarrow} \varepsilon^B_{\uparrow\downarrow}
 - n_{\uparrow\downarrow}  ( \varepsilon-\mathcal{E}^U_{\uparrow} )
 \right]
\frac{ U }{ \Lambda}
\end{eqnarray}
where
\begin{eqnarray}
\Lambda =  ( \varepsilon\!-\!\mathcal{E}^U_{\uparrow} )  ( \varepsilon\!-\!\mathcal{E}^U_{\downarrow} )
\!-\! (\!\varepsilon^B_{\uparrow\downarrow})^2,
\quad
\mathcal{E}^U_{\sigma}= \varepsilon_{\sigma}+U-i (\Gamma_{\sigma}+ \Gamma_{\bar\sigma} + \Gamma'_{\bar\sigma}),
\end{eqnarray}
$\varepsilon^B_{\uparrow\downarrow} = \varepsilon^B \sin\theta$, $\Gamma_{\sigma} = \sum_\ell\Gamma_{\ell\sigma} (\varepsilon)$, $\Gamma'_{\sigma} = \sum_\ell\Gamma_{\ell\sigma} (\varepsilon')$, $\widetilde{\Gamma}_{\sigma}=\sum_\ell
 \Gamma_{\ell\sigma} (\varepsilon) f^{FD}_\ell(\varepsilon)
+\Gamma_{\ell\sigma}(\varepsilon') f^{FD}_\ell(\varepsilon')$ and $\varepsilon'=2\epsilon_0+U-\varepsilon$.
While $G^R_{\sigma\sigma'}$ and $\langle\langle d_\downarrow, d^\dag_\uparrow \rangle\rangle$ can be solved generally from Eqs.~(\ref{eq:green_duu_ddu1}) and (\ref{eq:green_duu_ddu2}), practically for $U\gg eV \gg \{k_B T, \varepsilon_B, \Gamma \}$, solutions at only two energy windows are relevant in our case.
For $|\varepsilon-\varepsilon_0|\ll U$, i.e., type A impurity or for the lower energy level of type B impurity,
\begin{eqnarray}\label{eq:G_typeA}
G^R_{\uparrow\uparrow}
=
\frac
{
(1 - n_{\downarrow\downarrow})
(\varepsilon -\mathcal{E}_\downarrow )+
n_{\uparrow\downarrow}\varepsilon^B_{\uparrow\downarrow}
}
{
( \varepsilon -\mathcal{E}_\uparrow
)
(\varepsilon -\mathcal{E}_\downarrow
 )
-
(\varepsilon^B_{\uparrow\downarrow})^2
},
\qquad
G^R_{\downarrow\uparrow}
=
\frac
{n_{\uparrow\downarrow} (\varepsilon -\mathcal{E}_\uparrow )+
(1 - n_{\downarrow\downarrow}) \varepsilon^B_{\uparrow\downarrow}
}
{
( \varepsilon -\mathcal{E}_\uparrow
)
(\varepsilon -\mathcal{E}_\downarrow
 )
-
(\varepsilon^B_{\uparrow\downarrow})^2
},
\end{eqnarray}
where $\mathcal{E}_\sigma = \varepsilon_\sigma -i \Gamma_\sigma-i\tilde{\Gamma}_{\bar{\sigma}}$. For $|\varepsilon-(\epsilon_0+U)|\ll U$, i.e., for the higher energy level in type B impurity,
\begin{eqnarray}\label{eq:G_typeB}
G^R_{\uparrow\uparrow}
=
 \left[n_{\downarrow\downarrow} ( \varepsilon-\mathcal{E}^U_{\downarrow} -i \widetilde{\Gamma}_{\uparrow} ) -
n_{\uparrow \downarrow} \varepsilon^B_{\uparrow\downarrow} \right]
\frac
{\Lambda}{\Xi}
,
\qquad
G^R_{\downarrow\uparrow}
=
\left[ -  n_{\uparrow\downarrow} ( \varepsilon-\mathcal{E}^U_{\uparrow} -
 i \widetilde{\Gamma}_{\downarrow} )+
  n_{\downarrow\downarrow} \varepsilon^B_{\uparrow\downarrow}
\right]
\frac
{\Lambda}{\Xi},
\end{eqnarray}
where
\begin{eqnarray}
\Xi=
\left[ \Lambda-
 i \widetilde{\Gamma}_{\uparrow}( \varepsilon-\mathcal{E}^U_{\uparrow} )
\right]
 \left[ \Lambda-
 i \widetilde{\Gamma}_{\downarrow}( \varepsilon-\mathcal{E}^U_{\downarrow} )
\right]
+
  \widetilde{\Gamma}_{\uparrow}
  \widetilde{\Gamma}_{\downarrow}  (\varepsilon^B_{\uparrow\downarrow})^2.
\end{eqnarray}
$G^R_{\downarrow\downarrow}$ and $G^R_{\uparrow\downarrow}$ are obtained by reversing spin subscripts ($\uparrow\leftrightarrow \downarrow$) from the expressions of $G^R_{\uparrow\uparrow}$ and $G^R_{\downarrow\uparrow}$, respectively.

Assuming that $\Gamma(\varepsilon)$ varies slowly  on the scale of $\varepsilon_B$, integrals from Eq.~(\ref{eq:current_Langreth}) can be reduced to simple forms using Eqs.~(\ref{eq:G_typeA})-(\ref{eq:G_typeB}). If the Fermi levels obey $\mu_{l_1}-\mu_{l_2}\gg k_B T$ [$f^{FD}_{l_{1(2)}}=1(0)$ in the bias window], we have for type A
\begin{eqnarray}
\int d\varepsilon
\left(
 G^R_{\sigma' \sigma}  f^{FD}_{l_1}  +  \frac{1}{2}  G^<_{\sigma' \sigma}
\right)
&=&
\int^{\mu_{l_1}}_{-\infty} d\varepsilon\frac{1}{2}  G^>_{\sigma'\sigma}
=
-i \pi (1-n_{\uparrow\uparrow} - n_{\downarrow\downarrow})\delta_{\sigma\sigma'} ,
\label{eq:current_Langreth_A_in}
\\
\int d\varepsilon
\left(
 G^R_{\sigma'\sigma}  f^{FD}_{l_2}  +  \frac{1}{2}  G^<_{\sigma' \sigma}
\right)
&=&
\int^{\infty}_{-\infty} d\varepsilon\frac{1}{2}  G^<_{\sigma'\sigma}
=
i \pi  n_{\sigma \sigma'},
\label{eq:current_Langreth_A_out}
\end{eqnarray}
where we have used $G^R-G^A=G^>-G^<$, $\int^{\mu_{l_1}}_{\mu_{l_2}} d\varepsilon (G^R+G^A)=0$ as well as $\int^{\infty}_{\mu_{l_1}} d\varepsilon {\rm Im}G^R_{\sigma\sigma}=-\pi n_{\bar\sigma\bar\sigma}$ (so that $G^>_{\sigma\sigma}=-2i\pi n_{\bar\sigma\bar\sigma}$), and $\int^{\mu_{l_1}}_{\mu_{l_2}} d\varepsilon ({\rm Im}G^R_{\bar\sigma{\sigma}})=-\pi n_{\sigma \bar\sigma}$ (so that $\int^{\mu_{l_1}}_{\mu_{l_2}} d\varepsilon G^>_{\sigma,\bar{\sigma}}=0$). For type B, we have
\begin{eqnarray}
\int d\varepsilon
\left(
 G^R_{\sigma' \sigma}  f^{FD}_{l_1}  +  \frac{1}{2}  G^<_{\sigma' \sigma}
\right)
&=&
\int^{\infty}_{-\infty} d\varepsilon\frac{1}{2}  G^>_{\sigma'\sigma}
=
-i \pi (\delta_{\sigma\sigma'}-n_{\sigma\sigma'} ),
\label{eq:current_Langreth_B_in}
\\
\int d\varepsilon
\left(
 G^R_{\sigma'\sigma}  f^{FD}_{l_2}  +  \frac{1}{2}  G^<_{\sigma' \sigma}
\right)
&=&
\int^{\infty}_{\mu_{l_2}} d\varepsilon\frac{1}{2}  G^<_{\sigma'\sigma}
=
i \pi ( n_{\uparrow\uparrow} + n_{\downarrow\downarrow}-1)\delta_{\sigma\sigma'},
\label{eq:current_Langreth_B_out}
\end{eqnarray}
where we have used additionally $\int^{\mu_{l_1}}_{\mu_{l_2}} d\varepsilon G^>_{\bar{\sigma}\sigma}=2i\pi n_{\sigma {\bar\sigma}}$ [note that $\int d\varepsilon (G^>_{\bar\sigma{\sigma}}-G^<_{\bar\sigma{\sigma}}) \equiv -2i\pi\langle d^\dag_\sigma d_{\bar\sigma}+ d_{\bar\sigma} d^\dag_\sigma \rangle=0$]. Plugging Eqs.~(\ref{eq:current_Langreth_A_in})-(\ref{eq:current_Langreth_B_out}) and (\ref{eq:current_Langreth}) back into Eqs.~(\ref{eq:EOM_nd}) and (\ref{eq:EOM_dupdown}), we have Eqs.~(4) of the main text and the counterparts for spin injection via type B impurities as
\begin{eqnarray}
\hbar\dot n_{\sigma\sigma}
&=&
(1+\sigma p)\Gamma_{F} (1 -  n_{\sigma\sigma})
-\Gamma_{N}(  n_{\uparrow\uparrow}+  n_{\downarrow\downarrow} -1)
-2\varepsilon_B \sin\theta {\rm Im}( n_{\bar{\sigma}\sigma }),
\\
\hbar\dot n_{{\bar\sigma} \sigma}
&=&
-\Gamma_{F} n_{\bar\sigma\sigma }
+ i \varepsilon_B [2\bar{\sigma}\cos\theta n_{\bar{\sigma}\sigma}
+ \sin\theta (n_{\sigma\sigma}-  n_{\bar{\sigma}\bar{\sigma}} )] .
\end{eqnarray}

\noindent{\bf 3. Special cases of internal fields in Eq.~(7) of the main text}.

(I) \underline{$\mathcal{F}(B_i,\theta_i,\phi_i)$ with axial symmetry around the magnetization direction F or spherical symmetry}. For the axial symmetry case, $ \mathcal{F}(B_i,\theta_i,\phi_i)= \tfrac{1}{2\pi}\mathcal{F}(B_i,z)$ ($z\equiv\cos\theta_i$),
\begin{eqnarray}
I  & =&  I_0 \left[ 1 -  (1-l)p^2 \int^{\infty}_0 dB_i  \int^1_{-1} d z  \frac{ B^2_\|\mathcal{F}(B_i,z) }
{\sqrt{ Y^2_1-Y^2_2 }}\right],
\end{eqnarray}
where
\begin{eqnarray}
Y_1 &=& B^2_F+B^2_e+B^2_i +2 B_e B_i z \cos\theta_e -\alpha p^2 B_\|^2 ,
\nonumber\\
Y_2&=& 2B_eB_i\sin\theta_e \sqrt{1-z^2}.
\end{eqnarray}
For the hyperfine interaction with $E'$ defects, shown in Figure~2(a), we assumed spherical symmetry, $\mathcal{F}(B_i,\theta_i,\phi_i)= \tfrac{1}{4\pi}\delta (B_i - B_{E'})$, where $B_{E'}=420~G$ is the empirical amplitude of this interaction \cite{Griscom_PRB79}.

For the hyperfine interaction with $P_d$ interfacial defects, shown in Figure~2(b), we have followed the empirical findings of Ref.~\cite{Brower_APL83},
\begin{eqnarray}
 \mathcal{F}(B_i,\theta_i,\phi_i) = \frac{3}{8\pi}\frac{ (B_1^2 + B_2^2)\sin^2{\theta_i} + 2B_2^2\cos^2{\theta_i}}{B_1^2 + 2B_2^2} \frac{1}{\sqrt{2\pi}\sigma_0}\exp{\left\{ -\frac{1}{2}\left(\frac{B_i-B_0}{\sigma_0}\right)^2\right\}},
\end{eqnarray}
where $B_1$$\,$=$\,$160~G and $B_2$$\,$=$\,$90~G denote, respectively, the amplitudes of the out-of-plane and in-plane hyperfine-field components. The mean and variance of the normal distribution are $B_0 = (B_1 + 2B_2)/3$ and $\sigma_0$$\,$=$\,$20~G, respectively.

\vspace{3mm}

(II) \underline{$\mathcal{F}(B_i)$ due to exchange between nearest-neighbor impurities}. Assuming random distribution,  the probability of finding one impurity at distance $r$ while other impurities are located outside the sphere of radius $r$ is
\begin{eqnarray}
P(r)dr = 4\pi r^2 n_i dr \left(1-\frac{4\pi r^3 n_i}{3 N} \right)^{N-1}
= 4\pi r^2 n_i \exp\left(-\frac{4\pi r^3}{3}n_i\right) dr\,.
\end{eqnarray}
$n_i$ is the impurity spatial density and $N\rightarrow\infty$ is the number of impurities. We have used the identity $e=(1+1/x)^x|_{x\rightarrow\infty}$. The distribution probability $\mathcal{F}(B_i)$ can then be written as
\begin{eqnarray}
\mathcal{F}(B_i) = \left| \frac{P(r)}{d B_i(r)/dr}\right|.
\end{eqnarray}
We obtain $B_i(r)$ due to the exchange between nearest-neighbor impurities, $g\mu_B B_i=\langle J_{nn}\mathbf{S}\rangle$, by following the asymptotic approximation of the hydrogenic model used by Herring and Flicker \cite{Herring_PR64}
\begin{eqnarray}
J_{nn}(r) \approx 1.64 Ry \left(\frac{r}{a_B}\right)^{5/2} \exp\left(-\frac{2r}{a_B}\right),
\end{eqnarray}
where $Ry$ is Rydberg energy unit and $a_B$ is Bohr radius. The latter corresponds to the extension of the localized-electron wavefunction on the impurity. In Figure 2(c) of the main text we have used $4\pi n_i a_B^3/3 = 1/7^3$ and a $Ry$ unit of 200~meV that corresponds to $a_B = 0.44$~nm [$Ry = \hbar^2/(2m a_B^2)$]. They result in $n_i=8\times 10^{18}$ cm$^{-3}$. The results are similar for other choices that keep $n_i a_B^3$ constant. We integrate $B_i$ starting from 100~G, below which the impurities can be regarded as isolated ones where the hyperfine interaction overweighs the nearest neighbor exchange field.

\vspace{3mm}
\noindent{\bf 4. Effects from quantitative approximations applied in the main text.}

(I) \underline{Contribution from cotunneling}. Cotunneling is effective for impurities in the Coulomb blockade regime ($\varepsilon_0$ and $\varepsilon_0 +U$ are below and above the bias window respectively by more than the level broadening $\Gamma$ and thermal energy $k_B T$). From the outset it is a much smaller contribution to the tunnel current compared to the resonant tunneling. It is effectively a second-order perturbation process in which the electron tunnels via the intermediate virtual state of the impurity, $\varepsilon_0+U$. Being studied in detail in the  context of quantum dots more recently, the physical basis dates back to Kondo scattering setting, where the Kondo effect is from the third and higher order perturbation \cite{Kondo_PTP64, Appelbaum_PRL66}. By Schrieffer-Wolff transformation, it is clear that the Anderson impurity model [Eq.~(1)] leads to both spin independent and exchange tunneling amplitude $T_F T_N U/\varepsilon_0 (\varepsilon_0+U)$, where $T_{F(N)}$ is the coupling between the lead and impurity in Eq.~(1) of the main text \cite{Schrieffer_PR66}. This quantity should be compared to $T_F T_N /(\Gamma_F+\Gamma_N)$ in the resonant tunneling case. Compared to the Coulomb blockade regime, the resonant tunneling amplitude is much larger in general. Upon weak magnetic field, as shown in Eq.~(5) of the main text, a considerable MR effect occurs on the resonant current. Upon these considerations and under the usual condition of the tunnel energy window, $eV\gg \Gamma_F+\Gamma_N$, we can safely regard that the MR effect from resonant tunneling dominates in our study.

To  explicitly quantify the above statement, we focus on the strongest cotunneling region, i.e., either $\varepsilon_0$ or $\varepsilon_0+U$  is much closer to the bias window than the other level. We show the case that $\varepsilon_0$ is much closer which is similar to the other case. For $\mu_{l_1}>\mu_{l_2}$ and  $\varepsilon_0+U- \mu_{l_1}\gg \mu_{l_2}-\varepsilon_0 \gg \{\Gamma's, \varepsilon_B, k_B T\}$, i.e., in the Coulomb blockade regime, from Eqs.~(\ref{eq:green_duu_ddu1}) and (\ref{eq:green_duu_ddu2}) we have that for $\varepsilon< \mu_{l_1}$
\begin{eqnarray}
G^R_{\uparrow\uparrow}
=
\frac
{
(1 + \frac{U}{\varepsilon-\mathcal{E}^U_{\uparrow}} n_{\downarrow\downarrow})
(\varepsilon -\tilde{\mathcal{E}}_\downarrow )+
 \frac{U}{\varepsilon-\mathcal{E}^U_{\downarrow}} n_{\uparrow\downarrow}\varepsilon^B_{\uparrow\downarrow}
}
{
( \varepsilon -\tilde{\mathcal{E}}_\uparrow
)
(\varepsilon -\tilde{\mathcal{E}}_\downarrow
 )
-
(\varepsilon^B_{\uparrow\downarrow})^2
},
\qquad
G^R_{\downarrow\uparrow}
=
\frac
{-\frac{U}{\varepsilon-\mathcal{E}^U_{\downarrow}}n_{\uparrow\downarrow} (\varepsilon -\tilde{\mathcal{E}}_\uparrow )+
(1 + \frac{U}{\varepsilon-\mathcal{E}^U_{\uparrow}} n_{\downarrow\downarrow}) \varepsilon^B_{\uparrow\downarrow}
}
{
( \varepsilon -\tilde{\mathcal{E}}_\uparrow
)
(\varepsilon -\tilde{\mathcal{E}}_\downarrow
 )
-
(\varepsilon^B_{\uparrow\downarrow})^2
},
\end{eqnarray}
where $\tilde{\mathcal{E}}_\sigma = \varepsilon_\sigma -i \Gamma_\sigma+i  \frac{U}{\varepsilon-\mathcal{E}^U_{\sigma}}\tilde{\Gamma}_{\bar{\sigma}}$. Then the imaginary part of the integration in Eq.~(\ref{eq:current_Langreth}), relevant for Eqs.~(\ref{eq:EOM_nd}) and (\ref{eq:EOM_dupdown}), is carried out as
\begin{eqnarray}
{\rm Im}\int d\varepsilon
\left(
 G^R_{\sigma' \sigma}  f^{FD}_{l}  +  \frac{1}{2}  G^<_{\sigma' \sigma}
\right)
&=&
\int^{\mu_l}_{-\infty} d\varepsilon {\rm Im}  G^R_{\sigma' \sigma}    +  \frac{1}{2} \int  d\varepsilon {\rm Im} G^<_{\sigma' \sigma}
\nonumber \\
&\approx&
-\pi \left[(1-n_{\bar{\sigma}\bar{\sigma}})\delta_{\sigma,\sigma'} + n_{\sigma\bar{\sigma}}\delta_{\bar{\sigma},\sigma'} \right] \left[1-\frac{2(\Gamma_N+\Gamma_F)}{\mu_l-\varepsilon_0}\right]+\pi n_{\sigma\sigma'}.
\end{eqnarray}
Then Eqs.~(\ref{eq:EOM_nd}) and (\ref{eq:EOM_dupdown}) become
\begin{eqnarray}
 \hbar \dot  n_{\sigma\sigma}  & = & [\Gamma_N+(1\pm p)\Gamma_F] (1-n_{\uparrow\uparrow} - n_{\downarrow\downarrow}) -  2(\Gamma_N+\Gamma_F)\left[\frac{\Gamma_N}{\mu_N-\varepsilon_0} + \frac{(1\pm p)\Gamma_F}{\mu_F-\varepsilon_0}\right] (1 -n_{\bar{\sigma}\bar{\sigma}}) - 2 \varepsilon_B \sin{\theta} \text{Im}(n_{\bar{\sigma}\sigma}),\nonumber
 \\
 \hbar \dot n_{\sigma\bar{\sigma}}  & = &  i\varepsilon_B\left[  \sin{\theta}( n_{\bar\sigma\bar\sigma} - n_{\sigma\sigma}) \pm 2\cos{\theta} n_{\sigma\bar{\sigma}} \right] - 2(\Gamma_N+\Gamma_F)\left(\frac{\Gamma_N}{\mu_N-\varepsilon_0} + \frac{\Gamma_F}{\mu_F-\varepsilon_0}\right) n_{\sigma\bar{\sigma}}.
\end{eqnarray}
We solve $n_{\sigma\sigma'}$ at steady state and obtain the magnetic field modulated current
\begin{eqnarray}
i_{\rm cotun}& =&\frac{e}{\hbar} \left[2 \Gamma_N (1-n_{\uparrow\uparrow} - n_{\downarrow\downarrow}) - (2-n_{\uparrow\uparrow} - n_{\downarrow\downarrow})  2(\Gamma_N+\Gamma_F) \frac{\Gamma_N}{\mu_N-\varepsilon_0} \right]
\nonumber\\
&\approx& \frac{2e}{\hbar}\Gamma_N (\Gamma_N+\Gamma_F)
\left\{ \frac{2 \Gamma^2 \delta^2 [\delta^2 - p^2 \delta_F^2] + \varepsilon^2_B [2\delta^2 - p^2\delta_F^2 (1+\cos 2\theta)]}
{2\Gamma^2 \delta^2 [\Gamma \delta - p^2\Gamma_F\delta_F] + \varepsilon^2_B [2 \Gamma \delta - p^2\Gamma_F \delta_F (1+\cos 2\theta)]}
-\frac{1}{\mu_N-\varepsilon_0} \right\},
\end{eqnarray}
where $\Gamma=\Gamma_N+\Gamma_F$, $\delta=\delta_N+\delta_F$, and $\delta_{N(F)}= \Gamma_{N(F)}/[\mu_{N(F)}-\varepsilon_0]\ll 1$. We see that even for the strongest cotunneling regime ($\Gamma{\rm 's}\ll \mu_{l_2}-\varepsilon_0\ll U$), $i_{\rm cotun}$ is smaller by $\Gamma/(\mu_l-\varepsilon_0)$ than the resonant tunnel current [Eq.~(5) of the main text]. Correspondingly, the respective MR contribution from cotunneling is much smaller than that from resonant tunneling.  On top of this, we have a wide bias window $eV\gg \Gamma{\rm 's}$ within which the resonant tunneling is effective.

\vspace{3mm}

(II) \underline{Straightforward modification for finite temperature}. Eqs.~(4) and (5) are derived under the limit of $k_B T \ll eV$. The effect still qualitatively remains, while quantitatively reduced when $k_B T$ and $eV$ become comparable. The generalization is straightforward. With arbitrary Fermi distribution $f_{l}(\varepsilon) = {\exp[(\varepsilon-\mu_l)/k_B T]+1}^{-1}$ where $\mu_l$ is the Fermi level, one now treats injection and extraction together. For type A impurities, Eq.~(4) is extended as
\begin{eqnarray}
 \hbar \dot  n_{\sigma\sigma}  & = & \Gamma_N \left[ f_N(1-n_{\bar{\sigma}\bar{\sigma}}) -n_{\sigma\sigma}\right]  +(1 \pm p)\Gamma_F \left[ f_F(1-n_{\bar{\sigma}\bar{\sigma}}) - n_{\sigma\sigma}\right] - 2 \varepsilon_B \sin{\theta} \text{Im}(n_{\bar{\sigma}\sigma}),\nonumber
 \\
 \hbar \dot n_{\sigma\bar{\sigma}}  & = &  i\varepsilon_B\left[ \sin{\theta}( n_{\bar\sigma\bar\sigma} - n_{\sigma\sigma}) \pm 2\cos{\theta} n_{\sigma\bar{\sigma}} \right] + \left[\Gamma_N (f_N-1)+ \Gamma_F (f_F-1) \right] n_{\sigma\bar{\sigma}}.
\end{eqnarray}
From the steady state solution of $n_{\sigma\sigma}$ and from the electron-hole symmetry, one gets the resonant current 
\begin{eqnarray}\label{eq:current_temp}
i_A(f_N, f_F) & =&  -i_B(1-f_N, 1-f_F) =  \frac{2e}{\hbar} \frac{\Gamma_F \Gamma_N (f_N-f_F)}{\Gamma_N (1+f_N)+ \Gamma_F(1+f_F)} \frac{1 - p^2 \chi_{\scriptscriptstyle T}(\mathbf{B})}{1 - \alpha_{\scriptscriptstyle T} p^2 \chi_{\scriptscriptstyle T}(\mathbf{B})}, \nonumber \\
\chi_{\scriptscriptstyle T}(\mathbf{B})  &= &\frac{\Gamma_F(1-f_F)}{ \Gamma_F(1-f_F) +\Gamma_N(1-f_N)} \frac{[B_N(1-f_N) +B_F(1-f_F)]^2 + B^2 \cos^2\theta }{[B_N(1-f_N) +B_F(1-f_F)]^2 + B^2},  \\
\alpha_{\scriptscriptstyle T} &= & \frac{\Gamma_F (1+f_F)}{ \Gamma_N(1+f_N)+ \Gamma_F (1+f_F)}\,,\quad  B_{F(N)} = \frac{\Gamma_{F(N)}}{g\mu_B}.\nonumber
\end{eqnarray}
One can see that the MR of $i_A$ and $i_B$ comes from the $\mathbf{B}$ modulation of the expression $[1 - p^2 \chi_{\scriptscriptstyle T}(\mathbf{B})]/ [1 - \alpha_{\scriptscriptstyle T} p^2 \chi_{\scriptscriptstyle T}(\mathbf{B})]$ in the first equation of Eq.~(\ref{eq:current_temp}). It reaches the maximum 1 when $B\gg B_{N,F}$ and $\theta=\pi/2$ (total $\mathbf{B}$ out-of-plane), and reaches the minimum $(1-p^2)/(1-\alpha_{\scriptscriptstyle T} p^2)$ when $\theta=0$ (total $\mathbf{B}$ in-plane) or $\mathbf{B}=0$.

\vspace{3mm}

\noindent{\bf 5. More details on the effects of impurity locations, on the MR effect due to lateral bias distribution, and on the ultrashort spin lifetime in hole bands.}

(I) \underline{The impurity proximity to F or N leads and its effect on MR}. It is clear from Eq.~(5) that the MR effect is more effective when $\alpha$ is closer to 0 rather than 1. This corresponds to $\Gamma_N\gg \Gamma_F$, i.e., the impurity closer to N. This criterion can guide the F-I-N junction design that eliminates this MR effect. For example, direct Schottky contacts between magnetic metals and semiconductors may lead to measurable Hanle-type signals not overshadowed by the orders-of-magnitude larger MR effect \cite{Ando_APL11,Lou_PRL06,Chan_PRB09}. In this case, $\Gamma_N$ is associated with the Schottky barrier and the MR effect vanishes even if mid-gap states are present at the interface with F  (since $\Gamma_F \gg \Gamma_N$ still holds). On the other hand, many 3T measurements employ ferromagnet-oxide-semiconductor junctions. Even with plasma oxidation to eliminate the impurities inside the oxide barrier, the Schottky barrier impurities still produce MR effect due to our mechanism ($\Gamma_N\gg \Gamma_F$ due to the thicker oxide barrier).  As mentioned in the main text, the unique signatures observed for the dependence of tunnel resistance and MR on the oxide barrier thickness in Ref.~\cite{Sharma_PRB14} can be fully accounted by our mechanisms. In Ref.~\cite{Sharma_PRB14}, their control experiments with metal replacing semiconductor, and with tunneling off spin-polarized lead by insertion of non-magnetic layer also similarly agree with our theory.

\vspace{3mm}

(II) \underline{The MR effect due to lateral bias distribution under a floating contact}. Our mechanism also explains a series of `thermal spin injection' experiments (published in Ref.~\cite{Breton_Nat11} and elsewhere), where a similar orders-of-magnitude deviation exists from expected theoretical value if spin accumulation and relaxation is assumed to be the main cause. Although a floating F-I-N contact is used, (1) the lateral bias drop under the contact is  considerable ($\sim 0.1$ eV) and (2) the  tunnel MR is highly dependent on current direction (see Fig.~S4 of Ref.~\cite{Breton_Nat11}). These two aspects render our MR mechanism effective and dominant in their voltage measurement. When a large `heating' current is on \cite{Breton_Nat11}, there is a tunnel current crossing the floating junction at its one end and crossing with the opposite direction at its other end. With the highly current polarization dependent MR, one can show by an equivalent circuit that the voltage change for $V_{\rm Si}> V_{\rm FM}$ due to MR effect will pass on to their measured voltage change. A more or less 50 mV voltage across the F-I-N junction will produce the observed 0.13 mV MR signal under sufficient magnetic field, which can be indicated from Fig.~S4 of \cite{Breton_Nat11}. The full details will be reported elsewhere.

\vspace{3mm}

(III) \underline{The ultrashort spin lifetime in hole bands of semiconductors such as Si, Ge, and GaAs}.  It is well known the spin relaxation of holes is ultrafast in typical bulk semiconductors that are not subjected to large strain (see \cite{Zutic_RMP04} and references therein). Specifically, the spin relaxation rate of holes is governed by scattering between heavy and light holes. Due to the band degeneracy in the top of the valence band, the strong spin mixing of light-hole states renders this scattering ultrafast irrespective of the amplitude of the atomic spin-orbit coupling. The result is that spin and momentum relaxation rates of holes are in the same ballpark (typically sub picosecond regime). That the observed signals in F-I-N structures with $n$-type and $p$-type semiconductors have comparable amplitudes and widths ($\hbar/g \mu_B \Delta B \sim 0.1$~ns) strengthens the argument that the proposed MR effect corresponds to the measured signal.